\documentclass[aps,pra,showpacs,groupedaddress,twocolumn]{revtex4}

\usepackage{graphicx}
\usepackage{dcolumn}
\usepackage{bm}
\usepackage[usenames,dvipsnames]{color}
\usepackage{amsmath}
\usepackage[hypertex]{hyperref}
\usepackage{amsfonts}
\usepackage{amsthm}
\usepackage{amssymb}
\usepackage{mathrsfs}
\usepackage{subfigure}
\usepackage{ulem}

\begin{document}

\title{Quantum discord of ensemble of quantum states}

\author{Yao Yao}

\author{Jing-Zheng Huang}

\author{Xu-Bo Zou}
\email{xbz@ustc.edu.cn}

\author{Zhen-Qiang Yin}

\author{Wei Chen}

\author{Guang-Can Guo}

\author{Zheng-Fu Han}
\affiliation
 {Key Laboratory of Quantum Information, University of Science and Technology of China, Hefei, 230026,
 China}

\date{\today}

\begin{abstract}
We highlight an information-theoretic meaning of quantum discord as the gap
between \textit{the accessible information} and \textit{the Holevo bound} in the framework of
ensemble of quantum states. This complementary relationship implies that a large amount of pre-existing arguments about
the evaluation of quantum discord can be directly applied to the accessible information and vice versa.
For an ensemble of two pure qubit states, we show that one can evade the optimization problem with
the help of the Koashi-Winter relation. Further, for the general case (two mixed qubit states),
we recover the main results presented by Fuchs and Caves [Phys. Rev. Lett. \textbf{73}, 3047 (1994)],
but totally from the perspective of quantum discord.
Following this line of thought, we also investigate the geometric discord as an indicator
of quantumness of ensembles in detail. Finally, we give an example to elucidate the difference
between quantum discord and geometric discord with respect to optimal measurement strategies.
\end{abstract}


\maketitle

\textbf{Introduction.}
Since its introduction in 2001, the concept of quantum discord \cite{Ollivier2001,Henderson2001} has been gradually recognized
and used as an indicator of the quantumness of composite quantum systems. Most recently, it is arousing increasing interest
in its quantification and applications (for a recent review see \cite{Modi2011} and references therein). Quantum discord
originates from the inequivalence of two classically identical expressions of mutual information in the quantum realm.
For a composite bipartite system $\rho^{ab}$, the quantum mutual information is defined as
\begin{eqnarray}
I(\rho^{ab}):=S(\rho^{a})+S(\rho^{b})-S(\rho^{ab}),
\end{eqnarray}
where $S(\rho)=-tr\rho\log_2\rho$ is the von Neumann entropy, and $\rho^{a(b)}=tr_{b(a)}\rho^{ab}$ denote the reduced density operator
of subsystem A(B). On the other hand, consider performing a general measurement $\{\Pi^{b}_{k}\}$ on subsystem B.
An alternative version of the mutual information, proposed by Henderson and Vedral \cite{Henderson2001}, can be regarded
as a measure to quantify the purely classical part of correlations
\begin{align}
J_B(\rho^{ab}):&=S(\rho^{a})-S(\rho^{ab}|\{\Pi^{b}_{k}\}),\nonumber\\
&=S(\rho^{a})-\min_{\{\Pi^{b}_{k}\}}\sum_kp_kS(\rho^a_k),
\end{align}
with $p_k=tr\Pi^{b}_{k}\rho^{ab}$ and $\rho^a_k=tr_b\Pi^{b}_{k}\rho^{ab}/p_k$.
The information discrepancy between these two quantities is defined as the so called quantum discord
\begin{align}
D_B(\rho^{ab}):&=I(\rho^{ab})-J_B(\rho^{ab}),\nonumber\\
&=S(\rho^b)-S(\rho^{ab})+\min_{\{\Pi^{b}_{k}\}}\sum_kp_kS(\rho^a_k).
\end{align}
Note that the optimization problem is involved in the definition, so up to now we have only
obtained analytical results in some limited cases \cite{Luo2008,Ali2008,Chen2011}.

In this paper, we highlight an information-theoretic meaning of quantum discord as the information gap
between \textit{the accessible information} and \textit{the Holevo bound} in the framework of
ensemble of quantum states. The implication of this \textit{complementary} relationship is transparent and remarkable:
a great deal of pre-established arguments about the accessible information can be directly applied
to the (analytical) evaluation of quantum discord, even if there was no formal definition of quantum discord
at that time. As an accompaniment, we also investigate the geometric discord as a figure of merit
for characterizing quantum correlations of ensembles following the same line of thought. We hope that
this work can enlarge the scope of (analytical) evaluation of quantum discord and shed some new light
on the relation between quantum discord and quantum communication.

\textbf{Quantum discord as information gap.}
In the context of quantum ensembles, even a system so simple as one consisting of
only two nonorthogonal states can be surprisingly rich in physics \cite{Fuchs1998}. Now it has been realized
that the \textit{quantumness} of quantum ensembles can not be accounted for by considering only one specific
physical quantity. Therefore, a large amount of investigations have been devoted to the study of this topic
and several proposals have been made about how to classify and quantify the quantumness of ensembles
\cite{Fuchs1994,Fuchs1996a,Fuchs2003,Fuchs1996b,Horodecki2006,Horodecki2007,Luo2009,Luo2010,Luo2011,Zhu2011}.
Here we reveal a new information-theoretic meaning of quantum discord as the discrepancy
between \textit{the accessible information} and \textit{the Holevo bound} in the framework of
ensemble of quantum states. Assume that Alice prepares a quantum ensemble $\mathcal{E}=\{\lambda_i,\rho_i\}$, and then
Bob performs a POVM measurement $M=\{M_j\}$. The overall ensemble of states is $\rho=\sum_i\lambda_i\rho_i$.
The celebrated Holevo theorem declares that \cite{Holevo1973}
\begin{align}
H(A:B)\leq\chi(\mathcal{E}):=S(\rho)-\sum_i\lambda_iS(\rho_i),
\end{align}
where the Shannon mutual information $H(A:B)$ represents the classical mutual information between Ailce's preparation and Bob's
measurement outcome under the measurement $M$, $S(\rho)=-tr\rho\log_2\rho$ is the von Neumann entropy and $\chi(\mathcal{E})$
is the Holevo bound.

However, we observe that the quantum ensemble can be rephrased as a classical-quantum state
\begin{align}
\label{c-q}
\varrho_{\mathcal{E}}=\sum_i\lambda_i|i\rangle\langle i|\otimes\rho_i^{b},
\end{align}
Here $\{|i\rangle\}$ is an orthonormal basis of subsystem $A$.
With the help of the relation \cite{Nielsen}
\begin{align}
S\left(\sum_i\lambda_i|i\rangle\langle i|\otimes\rho_i^b\right)=H(\lambda_i)+\sum_i\lambda_iS(\rho_i^b),
\end{align}
It turns out that the \textit{quantum} mutual information of $\varrho(\mathcal{E})$ is equivalent to
the Holevo bound
\begin{align}
\label{quantum}
I[\varrho_{\mathcal{E}}]=S(\rho^b)-\sum_i\lambda_iS(\rho_i^b)=\chi(\mathcal{E}),
\end{align}
On the other hand, the (local) accessible information is defined as the maximum amount of \textit{classical}
mutual information that Bob can extract by measurement $M$
\begin{align}
\label{classical}
I_{acc}(\varrho_{\mathcal{E}})=\max_{M}I[M(\varrho_{\mathcal{E}})]=\max_{M}H(A:B),
\end{align}
From Eqs. (\ref{quantum}) and (\ref{classical}), it can be seen clearly that the \textit{information difference}
between the Holevo bound and accessible information is exactly the quantum discord, due to the original definition made by
Olivier and Zurek \cite{Ollivier2001}
\begin{align}
\label{discord}
I[\varrho_{\mathcal{E}}]-\max_{M}I[M(\varrho_{\mathcal{E}})]=\chi(\mathcal{E})-I_{acc}(\varrho_{\mathcal{E}})=D(\varrho_{\mathcal{E}}).
\end{align}
This remarkable relationship implies that given a quantum ensemble, quantum discord and the accessible information
are \textit{complementary} to each other, or more importantly, they share the \textit{same} optimal measurement strategy.
The above observation further suggests that a large amount of pre-existing arguments \cite{Davis1978,Peres1991,Hausladen1994,Jozsa1994,Sasaki1999}
(in fact most of which focus on seeking optimal measurements) about the accessible information can be directly
applied to the evaluation of quantum discord, even if at that time there was \textit{no} formal definition of quantum discord,
or in other words, their mathematical difficulties are equivalent.

In fact, this interpretation of quantum discord has already been noticed by Luo \textit{et al.} \cite{Luo2010}
(only a few examples were reported there).
Nevertheless, this important issue has never been further explored in the study of more general cases till now.
In the following, we investigate the quantum ensembles totally from the perspective of quantum discord
and revisit (and deepen) the previous results about the accessible information.
Since here we prefer to obtain analytical results, we concentrate on an ensemble of two states on two-dimensional
Hilbert spaces (i.e., qubit), the same as what was displayed in Ref. \cite{Fuchs1994}. However, this idea can be extended
to ensembles of more than two states and in high dimensions.

\textbf{Accessible information for ensemble of pure states.}
As an important application, we first demonstrate that for an ensemble of two pure states $|\phi_0\rangle$ and $|\phi_1\rangle$,
we can evade the optimization problem with the help of the Koashi-Winter relation \cite{Koashi2004}, to evaluate the
quantum discord of the bipartite state associated with the ensemble. Here the overall state can be written as
\begin{align}
\varrho^{ab}=\lambda_0|0\rangle\langle0|\otimes|\phi_0\rangle\langle\phi_0|+\lambda_1|1\rangle\langle1|\otimes|\phi_1\rangle\langle\phi_1|,
\end{align}
which can be purified to a tripartite pure state (qubit $C$ is an auxiliary system)
\begin{align}
|\Phi\rangle=\sqrt{\lambda_0}|0\rangle_a|\phi_0\rangle_b|0\rangle_c+\sqrt{\lambda_1}|1\rangle_a|\phi_1\rangle_b|1\rangle_c,
\end{align}
where $\varrho^{ab}=tr_c|\Phi\rangle\langle\Phi|$ and $\varrho^{ac}=tr_b|\Phi\rangle\langle\Phi|$.

The Koashi-Winter relation tells us that
\begin{align}
D^{\leftarrow}(\varrho^{ab})=E(\varrho^{ac})-S_{a\mid b},
\end{align}
where $D^{\leftarrow}(\varrho^{ab})$ denotes the quantum discord of $\varrho^{AB}$ with the subsystem $B$ measured,
$E(\varrho^{ac})$ is the entanglement of formation of $\varrho^{ac}$, and $S_{a|b}=S(\varrho^{ab})-S(\varrho^{b})$.
Note that $E(\varrho^{ac})$ is a monotonic function of concurrence $\mathcal{C}(\varrho^{ac})$
\begin{align}
E(\varrho^{ac})=h\left(\frac{1+\sqrt{1-\mathcal{C}(\varrho^{ac})^2}}{2}\right),
\end{align}
where $h(x)=-x\log_2x-(1-x)\log_2(1-x)$ is the binary entropy function.
In view of the density matrix of $\varrho^{ac}$, one can easily obtain the concurrence
\begin{align}
\mathcal{C}(\varrho^{ac})=2\sqrt{\lambda_0\lambda_1}\langle\phi_0|\phi_1\rangle,
\end{align}
Besides, for the density matrix $\varrho^{ab}$ we have
\begin{align}
S(\varrho^{ab})&=H(\{\lambda_i\})+\sum_{i=0,1}\lambda_iS(\rho_i)=h(\lambda_0),\\
S(\varrho^{b})&=-\lambda_{\pm}\log_2\lambda_{\pm},
\end{align}
with the eigenvalues of $\varrho^{B}=\lambda_0\rho_0+\lambda_1\rho_1$ being
$\lambda_{\pm}=\frac{1}{2}(1\pm\sqrt{1-4\lambda_0\lambda_1(1-\langle\phi_0|\phi_1\rangle^2)})$.
Hence, the quantum discord of this ensemble is analytically achieved
\begin{align}
D^{\leftarrow}(\varrho^{ab})=h\left(\frac{1+\sqrt{1-\mathcal{C}(\varrho^{ac})^2}}{2}\right)+h(\lambda_{+})-h(\lambda_0).
\end{align}
For the case $\lambda_0=\lambda_1=\frac{1}{2}$, we easily recover the results presented
in Ref. \cite{Fuchs1998,Levitin1995}, but totally from a different viewpoint.
Note that Fuchs's quantumness measure \cite{Fuchs1998} is naturally compatible
with the definition of quantum discord.

In addition to these technical points, we are more concerned with the optimal measurement
strategy for quantum discord. Later, we restrict our considerations to the more general case of
two mixed qubit states.

\textbf{Optimal strategy for quantum discord.}
In the Bloch representation, the two mixed states can be written as $\rho_0=\frac{1}{2}(\mathbf{1}+\vec{a}\cdot\vec{\sigma})$
and $\rho_1=\frac{1}{2}(\mathbf{1}+\vec{b}\cdot\vec{\sigma})$, where $a=|\vec{a}|\leq1$,
$b=|\vec{b}|\leq1$ and $\vec{\sigma}=(\sigma_x,\sigma_y,\sigma_z)$ being the
Pauli spin vector. To go through all possible one-qubit projective measurements,
we adopt the projectors  $\Pi_{\pm}=\frac{1}{2}(I\pm\vec{n}\cdot\vec{\sigma})$
with $n=|\vec{n}|=1$.

Accordingly, the post-measurement states $\rho^a_{\pm}=tr_b\Pi_{\pm}\varrho_{\mathcal{E}}\Pi_{\pm}/p_{\pm}$ are
\begin{equation}
\rho^a_{+}=
\left(\begin{array}{cc}
\lambda_0\frac{1+\vec{a}\cdot\vec{n}}{1+\vec{c}\cdot\vec{n}} & 0 \\
0 & \lambda_1\frac{1+\vec{b}\cdot\vec{n}}{1+\vec{c}\cdot\vec{n}}
\end{array}\right),
\end{equation}
\begin{equation}
\rho^a_{-}=
\left(\begin{array}{cc}
\lambda_0\frac{1-\vec{a}\cdot\vec{n}}{1-\vec{c}\cdot\vec{n}} & 0 \\
0 & \lambda_1\frac{1-\vec{b}\cdot\vec{n}}{1-\vec{c}\cdot\vec{n}}
\end{array}\right),
\end{equation}
where we introduce $\vec{c}=\lambda_0\vec{a}+\lambda_1\vec{b}$. The corresponding probabilities
$p_{\pm}=tr\Pi_{\pm}\varrho(\mathcal{E})\Pi_{\pm}$ are given by
\begin{align}
p_{\pm}&=\frac{1}{2}\lambda_0(1\pm\vec{a}\cdot\vec{n})+\frac{1}{2}\lambda_1(1\pm\vec{b}\cdot\vec{n}),\nonumber\\
&=\frac{1}{2}(1\pm\vec{c}\cdot\vec{n}),
\end{align}

The key point of evaluating quantum discord is to search the minimum value of
the conditional quantum entropy
\begin{align}
\label{conditional}
\mathcal{S}(\vec{n})=S(A|\Pi)=p_+S(\rho^A_{+})+p_-S(\rho^A_{-}),
\end{align}
The optimal projector for Eq. (\ref{conditional}) can be found by varying it with respect to all unit vectors $\vec{n}$,
that is, by setting $\delta\mathcal{S}(\vec{n})=0$. The resulting condition for the optimal $\vec{n}$ is
\begin{align}
\label{condition1}
\left[(\lambda_0\log_2\vartheta_0)\vec{a}+(\lambda_1\log_2\vartheta_1)\vec{b}\right]\cdot\delta\vec{n}=0,
\end{align}
where we define
\begin{align}
\vartheta_0=\frac{(1+\vec{a}\cdot\vec{n})(1-\vec{c}\cdot\vec{n})}{(1-\vec{a}\cdot\vec{n})(1+\vec{c}\cdot\vec{n})},\nonumber\\
\vartheta_1=\frac{(1+\vec{b}\cdot\vec{n})(1-\vec{c}\cdot\vec{n})}{(1-\vec{b}\cdot\vec{n})(1+\vec{c}\cdot\vec{n})},
\end{align}
Here we notice that an infinitesimal variation of the \textit{unit} vector is an infinitesimal rotation, i.e.
$\delta\vec{n}=\vec{\epsilon}\times\vec{n}$, where $\vec{\epsilon}$ is an arbitrary infinitesimal vector.
This indicates that $\delta\vec{n}$ is perpendicular to $\vec{n}$. Therefore, if we divide a vector $\vec{a}$
into two parts $\vec{a}_\perp$ and $\vec{a}_\parallel$ (here the subscripts $\perp$ and $\parallel$ are with
respect to $\vec{n}$), only the $\vec{a}_\perp$ ($\vec{b}_\perp$) part can survive in Eq. (\ref{condition1}).
Hence, Eq. (\ref{condition1}) becomes
\begin{align}
\left[(\lambda_0\log_2\vartheta_0)\vec{a}_\perp+(\lambda_1\log_2\vartheta_1)\vec{b}_\perp\right]\cdot\delta\vec{n}=0,
\end{align}
where $\vec{a}_\perp=\vec{a}-(\vec{a}\cdot\vec{n})\vec{n}$ and $\vec{b}_\perp=\vec{b}-(\vec{b}\cdot\vec{n})\vec{n}$.
This equation further suggests that our final condition is
\begin{align}
(\lambda_0\log_2\vartheta_0)\vec{a}_\perp+(\lambda_1\log_2\vartheta_1)\vec{b}_\perp=\vec{0},
\end{align}
This is \textit{exactly} the same condition that should be satisfied by the optimal $\vec{n}$
associated with the accessible information \cite{Fuchs1994} (see also \cite{Fuchs1996a}). Moreover,
if we let
\begin{eqnarray}
\left\{\begin{array}{cc}
\vartheta_0=\vartheta_1^{-1}, \\
\lambda_0\vec{a}_\perp+\lambda_1\vec{b}_\perp=\vec{0}.
\end{array}\right.
\label{final1}
\end{eqnarray}
one can precisely recover the three cases raised in Ref. \cite{Fuchs1994} which can be analytically solved,
and this derivation in turn enriches the instance of analytical exploration of quantum discord.

\textbf{Optimal strategy for geometric discord.}
Following the above line of thought, we turn to investigate the optimal measurement strategy
for geometric discord, which was introduced as a geometrical way of quantifying quantum discord \cite{Dakic2010}
\begin{align}
D_{G}(\rho):=\min_{\chi\in\Omega}\|\rho^{ab}-\chi^{ab}\|^2,
\end{align}
where $\Omega$ denotes the set of ($B$-side) zero-discord states and $\|\rho-\chi\|^2=Tr(\rho-\chi)^2$ is the square of
Hilbert-Schmidt norm. It is worth mentioning that, Luo and Fu presented a simplified but equivalent version of
the geometric discord \cite{Luo2010a}
\begin{align}
\label{geometric}
\mathcal{D}_{G}(\rho^{ab})=\min_{\Pi^b}||\rho^{ab}-\Pi^b(\rho^{ab})||^2,
\end{align}
where the minimum is over all von Neumann measurements $\Pi^b=\{\Pi^b_k\}$ on subsystem $B$.

Suppose $\Pi=\{\Pi_i\}$ is a complete set of orthogonal projectors and $\rho'=\Pi(\rho)=\sum_i\Pi_i\rho\Pi_i$,
then we have the identity \cite{Nielsen}
\begin{align}
-tr\rho\log_2\rho'=S(\rho'),
\end{align}
Therefore, Eq. (\ref{geometric}) can be further simplified to
\begin{align}
\label{simplified}
D_{G}(\rho)=tr(\rho^2)-\max_{\Pi^b}tr[(\Pi^b(\rho))^2],
\end{align}
which can be viewed as the minimum \textit{purity} deficit. Recall that
the overall state of our ensemble is $\varrho_{\mathcal{E}}=\lambda_0|0\rangle\langle0|\otimes\rho_0+\lambda_1|1\rangle\langle1|\otimes\rho_1$,
and the first term of Eq. (\ref{simplified}) can be easily obtained
\begin{align}
\label{term1}
tr(\varrho_{\mathcal{E}}^2)=\frac{1}{2}\left[\lambda_0^2(1+a^2)+\lambda_1^2(1+b^2)\right],
\end{align}
To arrive at the maximum valve of the second term, we also need to optimize over all
von Neumann measurements $\Pi^b=\{\Pi_{\pm}\}$ as we did in the previous section.
We observe that
\begin{align}
[\Pi^b(\varrho_{\mathcal{E}})]^2 &= \left[\sum_{i=\pm}\Pi_i\varrho_{\mathcal{E}}\Pi_i\right]^2,\nonumber\\
&= \lambda_0^2|0\rangle\langle0|\otimes(\Pi_+\rho_0\Pi_+)^2\nonumber\\
& +\lambda_1^2|1\rangle\langle1|\otimes(\Pi_+\rho_1\Pi_+)^2\nonumber\\
& +\lambda_0^2|0\rangle\langle0|\otimes(\Pi_-\rho_0\Pi_-)^2\nonumber\\
& +\lambda_1^2|1\rangle\langle1|\otimes(\Pi_-\rho_1\Pi_-)^2,
\end{align}
Note that we have the identities
\begin{align}
\Pi_\pm\rho_0\Pi_\pm=(1\pm\vec{a}\cdot\vec{n})\Pi_\pm,\nonumber\\
\Pi_\pm\rho_1\Pi_\pm=(1\pm\vec{b}\cdot\vec{n})\Pi_\pm,
\end{align}
Thus, we have
\begin{align}
\label{term2}
tr[(\Pi^b(\varrho_{\mathcal{E}}))^2]
=&\frac{1}{2}\lambda_0^2[1+(\vec{a}\cdot\vec{n})^2)]\nonumber\\
&+\frac{1}{2}\lambda_1^2[1+(\vec{b}\cdot\vec{n})^2)],
\end{align}

By setting $\delta D_{G}(\vec{n})=0$, the condition for optimal $\vec{n}$ is
\begin{align}
\left[\lambda_0^2(\vec{a}\cdot\vec{n})\vec{a}+\lambda_1^2(\vec{b}\cdot\vec{n})\vec{b}\right]\cdot\delta\vec{n}=0,
\end{align}
The same reasoning (with respect to Eq. (\ref{condition1})) leads us to
\begin{align}
\label{condition2}
\lambda_0^2(\vec{a}\cdot\vec{n})\vec{a}_\perp+\lambda_1^2(\vec{b}\cdot\vec{n})\vec{b}_\perp=\vec{0},
\end{align}
To satisfy Eq. (\ref{condition2}), two possible choices may be
\begin{eqnarray}
\left\{\begin{array}{cc}
\vec{a}\cdot\vec{n}=\pm\vec{b}\cdot\vec{n}, \\
\lambda_0^2\vec{a}_\perp\pm\lambda_1^2\vec{b}_\perp=\vec{0}.
\end{array}\right.
\label{choice}
\end{eqnarray}
Later we will show that in some analytical cases the interchange between this two choices
makes the optimal strategy of geometric discord very different form that of original discord.

\begin{figure}[htbp]
\begin{center}
\includegraphics[width=.40\textwidth]{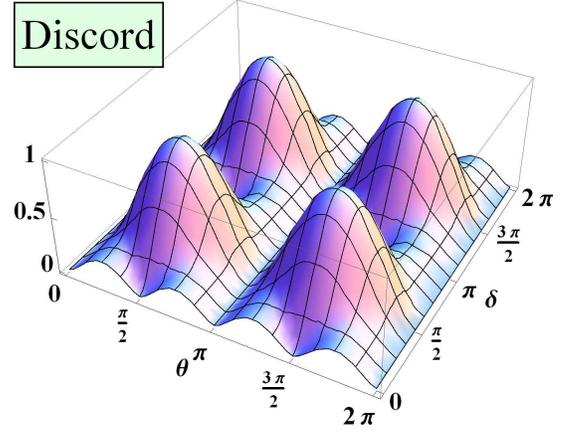} {}
\end{center}
\caption{(Color online) The rough quantum discord $\widetilde{D}(\theta,\delta)$ (before optimization) as a function of $\theta$ and $\delta$.
Here the Bloch vector of von Neumann measurement is $\vec{n}=(\cos\delta,0,\sin\delta)$.}
\label{f1}
\end{figure}

\textbf{An explicit example and comparison.}
Finally, to illustrate the above arguments about measurement strategies, we focus on
a specified ensemble of two pure states with equal probabilities (i.e., $\lambda_0=\lambda_1=\frac{1}{2}$).
Here we define
\begin{align}
|\phi_0\rangle=\cos\frac{\theta}{2}|0\rangle+\sin\frac{\theta}{2}|1\rangle,\nonumber\\
|\phi_1\rangle=\cos\frac{\theta}{2}|0\rangle-\sin\frac{\theta}{2}|1\rangle,
\end{align}
where $\langle\phi_0|\phi_1\rangle=\cos\theta$ and the corresponding Bloch vectors are
$\vec{a}=(\sin\theta,0,\cos\theta)$, $\vec{b}=(-\sin\theta,0,\cos\theta)$ respectively.
From Eq. (\ref{final1}), the requirement of optimal $\vec{n}$ for quantum discord is
equivalent to $\vec{n}\propto\vec{a}-\vec{b}$, which means that the direction of $\vec{n}$
is the bisectrix of the angle between $\vec{a}$ and $\vec{b}$. More precisely, $\Pi=\{\frac{1}{2}(I\pm\sigma_x)\}$
is the optimal observable for quantum discord. Actually, in a more visualizable way,
we can plot the (non-optimized) quantum discord $\widetilde{D}(\theta,\delta)$ as a function of $\theta$ and $\delta$ (here $\vec{n}=(\cos\delta,0,\sin\delta))$,
employing the algorithm proposed in our previous work \cite{Yao2012}. Form Figure \ref{f1}, it can be easily seen that
we can choose $\delta=0$ or $\pi$ to achieve quantum discord, for \textit{every} value of $\theta$, in other words,
\textit{irrespective} of the value of $\theta$, which is consistent with the above analysis. Consequently, the quantum discord
reads
\begin{align}
D(\varrho_{\mathcal{E}})=h\left(\frac{1+\sin\theta}{2}\right)+h\left(\frac{1+\cos\theta}{2}\right)-1,
\end{align}

On the other hand, one can analytically obtain the expression of the geometric discord from Eqs. (\ref{term1}) and (\ref{term2}) as
\begin{align}
D_G(\varrho_{\mathcal{E}})=&\frac{1}{4}-\max_\delta\frac{1}{4}(\sin^2\theta+\cos2\theta\sin^2\delta),\nonumber\\
=&\left\{\begin{array}{cc}
\frac{1}{4}\sin^2\theta, & \mbox{ if } \, \theta\in[0,\frac{\pi}{4})\cup(\frac{3\pi}{4},\pi]\\
\frac{1}{4}\cos^2\theta, & \mbox{ if } \, \theta\in[\frac{\pi}{4},\frac{3\pi}{4}]
\end{array}\right.\nonumber\\
=&\frac{1}{8}(1-|\cos2\theta|).
\end{align}
The calculation implies that when $\theta\in[0,\frac{\pi}{4})\cup(\frac{3\pi}{4},\pi]$, the optimal
measurement strategy for geometric discord is $\vec{n}=(0,0,1)$; however, when
$\theta\in[\frac{\pi}{4},\frac{3\pi}{4}]$, the relevant optimal measurement is $\vec{n}=(1,0,0)$.
In sharp contrast to the situation of quantum discord, the optimal strategy of geometric discord
\textit{depends} on the angle between the two pure states. Moreover, the transition of
this two kinds of optimal measurements just corresponds to the two possible choices in Eq. (\ref{choice}).

\begin{figure}[htbp]
\begin{center}
\includegraphics[width=.40\textwidth]{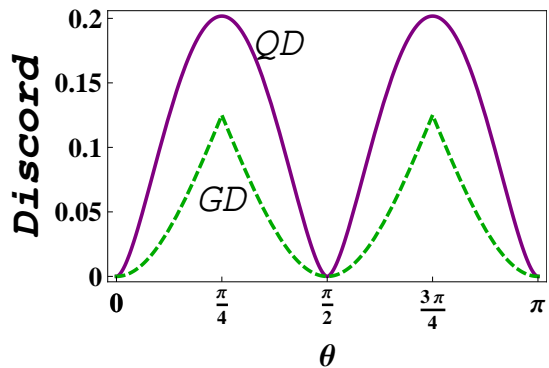} {}
\end{center}
\caption{(Color online) The comparison between quantum discord (purple solid line) and geometric discord (green
dashed line).}
\label{f2}
\end{figure}

As a comparison, we plot quantum discord and geometric discord for the ensembles together in Figure \ref{f2}.
We observe that geometric discord behaves monotonically
with respect to quantum discord and it indicates that in
this case geometric discord can also be viewed as a faithful measure of
quantumness of quantum ensembles, in spite of the transition of the optimal measurement strategies.

\textbf{Conclusion.} We highlight a significant information-theoretic meaning of quantum discord as the gap
between the accessible information and the Holevo bound in the context of quantum ensembles. This remarkable
relationship indicates that quantum discord and the accessible information share the \textit{same} optimal measurement strategy
and a large amount of pre-existing arguments about the evaluation of quantum discord can be directly applied to
the accessible information and vice versa. We have analytically
obtained the optimal measurement strategies of quantum discord and geometric discord for an ensemble of
two mixed (two-dimensional) states, which easily recover the results in Ref. \cite{Fuchs1994}. We
emphasize that this interpretation can be generalized to more general cases.
For instance, our analysis of geometric discord can be directly extended to ensembles of more that
two states. These results build a new bridge between quantum discord and quantum communication
and we hope that our attempt can attract more attention to this direction and enlarge the scope of
analytical evaluation of quantum discord.

\begin{acknowledgments}
Y.Y. wishes to thank Prof. Shunlong Luo for his helpful comments and valuable suggestions.
This work was supported by the National Basic Research Program of China (Grants No. 2011CBA00200 and No. 2011CB921200),
National Natural Science Foundation of China (Grants No. 60921091 and No. 61101137).
\end{acknowledgments}


\end{document}